\documentclass[a4paper,11pt,reqno]{amsart}
\pdfoutput=1

\usepackage{a4}
\usepackage{amsthm}
\usepackage[T1]{fontenc}
\usepackage[utf8]{inputenc}
\usepackage[british]{babel}
\usepackage{todonotes}
\usepackage[unicode,psdextra]{hyperref}
\usepackage{tikz}
\usetikzlibrary{decorations.pathmorphing,shapes}
\usetikzlibrary{arrows.meta,calc}
\def\centerarc[#1](#2)(#3:#4:#5)
    { \draw[#1] ($(#2)+({#5*cos(#3)},{#5*sin(#3)})$) arc (#3:#4:#5); }

\makeatletter
\@addtoreset{definition}{section}
\@addtoreset{equation}{section}
\makeatother

\makeatletter
\let\@wraptoccontribs\wraptoccontribs
\makeatother

\allowdisplaybreaks[4]

\begin{document}

\title[Quantum Field Theory Between Rigor and Pragmatism]{Quantum Field Theory Between Rigor and Pragmatism}

\author[J. Branahl]{Johannes Branahl\textsuperscript{1}}
 
\address{\textsuperscript{1}Philosophisches Seminar der
Universit\"at M\"unster \hfill \newline
Domplatz  23, 48143 M\"unster, Germany \hfill \newline
{\itshape e-mail:} \normalfont
\texttt{j\_bran33@uni-muenster.de}}

\begin{abstract}
Quantum Field Theory (QFT), the foundational framework of particle physics, has long existed in a state of tension between empirical success and mathematical rigor. Conventional QFT (CQFT), which underpins the Standard Model, offers unparalleled predictive accuracy but relies on inconsistent and ad hoc methods. In contrast, axiomatic QFT (AxQFT) aspires to a consistent, mathematically rigorous foundation, yet lacks empirical applicability. This paper introduces the heuristic vehicle of a “realism quotient”  to model and navigate this tension, framing it as a dynamic balance between realism and instrumentalism in the mathematical structures of physical theories. By reconstructing the historical development of QFT and extrapolating its trajectory, the paper offers both a descriptive account of theoretical progress and a normative proposal: that CQFT and AxQFT must converge to address selected challenges of physics beyond the Standard Model. The model also contributes to broader debates in scientific realism, offering a structured framework for understanding the interplay between empirical adequacy and conceptual robustness and mathematical rigour in foundational physics.
\end{abstract}

\keywords{(Structural) Realism, Instrumentalism, Empirical Adequacy, Fundamental Physics, Beyond the Standard Model}

\maketitle
\markboth{\hfill\textsc\shortauthors}{\textsc{{Quantum Field Theory Between Rigor and Pragmatism}\hfill}}

\section{Introduction}

For 70 years, the research program of quantum field theory —the theory of elementary particles—has been in a peculiar, yet highly intriguing state making it particularly interesting for philosophy of science. Conventional quantum field theory (CQFT) was decisive in constructing the Standard Model of particle physics, one of the great success stories of modern science, which describes the behavior of the smallest particles with unprecedented precision. However, the Standard Model did not resolve all problems—fundamental questions about the nature of subatomic and cosmic reality remain unanswered. Since the 1970s, the research program of BSM physics (beyond the Standard Model) has sought to formulate new theories, utilizing CQFT as its mathematical foundation. None of these numerous theories were empirically supported to this day, and many can be considered permanently ruled out. Two aspects are often cited as contributions to this crisis: the questionable methods of theory development and evaluation utilized in QFT, often based on metatheoretical criteria, and the technically limited scope of accelerator experiments. However, a third potential component of this crisis seems to be overlooked: the immature mathematics of CQFT, whose inconsistencies and rigor issues have been known since the early days of particle physics (e.g., \cite{Dyson1952}, \cite{Landau1954}, \cite{Haag1955}, \cite{Wightman1956}). The Standard Model required many pragmatic concessions, legitimized by its exceptional empirical adequacy, which may serve as a prime example of instrumentalist success. But could a revision of CQFT fuel progress beyond the Standard Model in the long term?

Complementary to the CQFT of BSM physics, the research program of axiomatic QFT (AxQFT) seeks to create a mathematically rigorous version of the hitherto inconsistent framework of particle physics. Despite remarkable structural results and the replication of important theorems, a rigorous formulation of a QFT in four dimensions remains absent (one of the seven Millennium Problems of mathematics). Thus, at least for progress in BSM physics, this research program is initially unusable. An underlying, often unspoken assumption permitting ontological insights in AxQFT is the Mathematical Universe Hypothesis (MUH, \cite{Tegmark2008}). It embodies a scientific realism regarding the mathematical structures of QFT and demands utmost rigor—inconsistencies must be entirely eliminated, as the structures and laws of our universe seems to exhibit no internal contradictions. Hence, this program is preferred for addressing questions about the ontological nature of elementary particles in philosophy rather than for predictions in practical accelerator experiments. Without the empirical adequacy of axiomatically constructed models, however, one can essentially only describe the attitude of the program’s proponents as idealistic, not realistic as the MUH suggests. But could a revision of AxQFT fuel progress beyond the Standard Model in the long term?
\\ \\
The two antipodal programs, AxQFT and CQFT, seemingly share a commonality in that both fail due to their radical positions regarding the significance of mathematical rigor for empirical adequacy. But how much rigor is necessary for an empirically adequate foundational physics—for the penultimate questions (quantum gravity is excluded) that BSM physics aims to explore? How closely must the mathematics used align with, or even correspond to the actual structures of this world? To answer these questions, this article develops a new refined model as a compromise solution for the intense debate on realism and instrumentalism, ongoing since the foundational works of the 1980s (e.g., \cite{vanFraassen1980}, \cite{Laudan1981}, \cite{Putnam1981}, \cite{Worrall1989}), aiming to bring the two QFT programs closer together. This model is on one hand intended to contribute to general philosophy of science. At the same time, it integrates into the specific philosophical debate about QFT, offering a conciliatory perspective in the Fraser-Wallace dispute regarding the two divergent QFT programs (e.g., \cite{Fraser2008}, \cite{Fraser2009}, \cite{Fraser2011} vs. \cite{Wallace2006}, \cite{Wallace2010}), which has recently gained momentum (e.g., Williams 2019, J.\,Fraser 2020, Rivat 2021, Dougherty 2023) and is becoming increasingly nuanced. 
The proposed model consists of a \textit{descriptive part}, which rationally reconstructs the historical evolution of modern foundational physics through the lens of a dynamic realism-instrumentalism quotient, and a \textit{normative part}, which provides guidance on balancing these elements in quantum field theory to address challenges in BSM physics. We formulate these two central results: 
\\ \\ 
 \textbf{Descriptive retrospection:} The history of modern foundational physics can be modeled as a process in which theories evolve with a steadily increasing ratio of realist to instrumentalist mathematical elements, required for empirical adequacy for each era of experimental precision. This "realism quotient" $R$ reflects the extent to which mathematical structures in a theory are assumed to correspond to underlying physical reality, as opposed to serving merely as tools for empirical prediction. $R$ was regularly adjusted in response to the increasing experimental resolution, which necessitates recalibration of theoretical frameworks and often provokes radical changes (given the incommensurable transition from instrumental to realistic mathematical elements). 

This historical perspective aligns with but extends the idea of convergent realism, emphasizing not just a general trend towards realism but the continuous interplay between realism and instrumentalism as a driver of theoretical evolution. Applied to BSM physics, the model gives concrete guidance for future theory development by extrapolating the trajectory of $R$:
\\ \\
 \textbf{Normative prospection:} CQFT and AxQFT fail due to their overly instrumentalist or overly realist stances (too many pragmatic compromises or overly high axiomatic demands) and should approach each other to achieve the balance required for a selection of open questions in BSM physics in terms of the rational reconstrution from above. 
\\ \\ 
A core message of this article is thus the necessary balancing act between required mathematical rigor and permissible (or inavoidable) pragmatism on the path towards a fundamental theory of physics. Does the empirical adequacy of the Standard Model, whose theoretical predictions rely on workarounds and inconsistent methods, justify this pragmatism? Or does the engagement with the fundamental constituents of the cosmos demand a stable mathematical foundation, as the world is increasingly revealed as mathematical structure in the abstractness of today’s particle physics? To develop and subsequently justify the model presented above, we proceed as follows: After some general remarks on scientific realism and instrumentalism, particularly on MUH and structural realism (Chapter 2), we position QFT within this field of tension (Chapter 3). Chapter 4 first constructs the mentioned compromise solution and then justifies it using historical examples from particle physics (Fermi’s theory of weak interaction, Gell-Mann's Eightfold Way, Heisenberg’s theory of everything, and Connes' non-commutative QFT), supplemented by practical guidelines for reconciling CQFT and AxQFT. Finally, in the concluding Chapter 5, we assess the article's contribution to philosophy of science in general and to the evaluation of the crisis in BSM physics in particular.  This article is intended for both philosophers of science and theoretical physicists concerned with foundational issues. While it draws on technical and historical insights from quantum field theory, it also engages with conceptual questions of realism, instrumentalism, and theory development. As such, it aims to offer a common conceptual framework—accessible yet analytically robust—that speaks across disciplinary lines.

\section{Realism and Instrumentalism in a Mathematical Universe}

\textit{Are our theories about the world true?} The dispute between scientific realism and instrumentalism is a central discourse in the philosophy of science. According to the realist perspective, scientific theories aim to provide a true\footnote{There are various subtleties differentiating the concepts of scientific realism present in the literature. We follow the definition rolled out in, among others, (\cite{vanFraassen1980}, \cite{Lyons2005}).} description of things in the world, including unobservable entities that exist independently of our theories about them. Scientific theories are intended to uncover the true nature of these entities. Instrumentalism, on the other hand, views scientific theories not as true or false\footnote{Again, given the various definitions of "instrumentalism" in the literature, we focus on the traditional form which is characterized by \cite{SEP2017} as follows: "statements involving (unobservable entities) are not even candidates for truth or falsity".} but as more or less useful models for predicting and explaining phenomena. Theoretical entities are treated as useful fictions that help make predictions and explain experiments without asserting their actual existence. Instrumentalism thus focuses on the practical applicability and success of theories, rather than their ontological truth. Consequently, scientific theories are merely classification schemes that make no ontological claims about the universe. The goals of science, as seen by realists and instrumentalists, could not be more different: for the former, it is the discovery and description of the true nature of the world, while for the latter, it is the development of pragmatic tools for predicting and explaining phenomena.

A preliminary approach to the current issues in QFT is offered by Worrall \cite{Worrall1989}. He introduced the concept of \textit{structural realism}, which seeks to combine the strengths of realism and instrumentalism. This position asserts that the mathematical (structural) aspects of theories are often retained even when their ontological assumptions are proven false (thus being a kind of selective realism). For example, the mathematical structure that gravitational strength decreases with the square of the distance—while not merely an instrumental stipulation in Einstein’s General Theory of Relativity—was recognized as a necessary consequence of the geometric nature of his theory and thus identified as a structural truth about gravity. According to Worrall, the mathematical equations of a theory may survive in its successor theory, suggesting that earlier theories could already capture structural truths about the world, even if they were ontologically mistaken. This supports the notion that there is structural continuity in science. Structural realism thus explains the success of scientific theories without assuming that their ontological claims about the world are correct. Worrall's ideas, however, can be sharpened by explicitly incorporating ontology. A central question must be addressed before the general debate between realism and instrumentalism can be concretely applied to quantum field theory: What is real? For the purposes of this work, the provisional answer to this fundamental question is: mathematics—at least within the confines of foundational physics. This answer has been extensively discussed in philosophy and physics. We will recapitulate its concrete formulation in the \textit{mathematical universe hypothesis (MUH)} and its implications for the realism debate described above before explaining its particular persuasiveness in the context of quantum field theory. 

The MUH, a provocative idea proposed by Tegmark \cite{Tegmark2008}, posits that the physical universe is a mathematical structure. Tegmark argues that if a mathematical structure can fully explain physical observations, then that structure should be regarded as reality itself\footnote{Thus, truth and reality become two sides of the same coin: Truth in foundational physics is understood as a property of theories (mathematical statements or propositions) that accurately reflect or correspond to reality. Reality, on the other hand, refers to the objective existence of things independent of our theories about them. According to Tegmark, \textit{some} mathematical truths are indistinguishable from the reality of our universe, collapsing the distinction between which theories are "true" in a logical sense and which entities described by theories are "real" in an ontological sense. For the purposes of this article, we leave the more sophisticated discussion about \textit{all} mathematical structures aside, being present in a hierarchy of parallel universes.}.  Unlike Worrall, a theory that successfully employs this mathematical structure cannot be ontologically mistaken. Tegmark's MUH stands in a long tradition of mathematical natural philosophy, from the Pythagoreans ("panta arithmos") through Galilei to modern physics, supported by the concept of \textit{ontic structural realim} (\cite{Ladyman1998}, \cite{Chakravartty2012}), extending Worrall's work \cite{Worrall1989} beyond mere \textit{epistemic} statements, where nothing can be said about the nature of these entities. The MUH radicalizes this tradition by suggesting that mathematical structure is not merely the language in which the laws of nature are written but the laws of nature themselves. If Tegmark is correct, the search for a "Theory of Everything" (TOE) is a quest for the underlying mathematical structure of our universe. This would represent the ultimate triumph of realism in foundational physics. One thing, however, is entirely clear: The universe, as a consistent mathematical structure, does not permit the coexistence of incompatible frameworks being still partly present in today’s research. Thus, only a subset of the mathematical structures utilized in today's theories can plausibly be considered real, as primarily existing entities in the universe. The remainder must be interpreted to be preliminary instrumentalist structures, perhaps to be discarded one day. 

The MUH does not resolve the ongoing debate between scientific realism and instrumentalism but at least refines the question about the essence of real entities. From a realist perspective, mathematical structures are not just tools or models for describing the physical world but the fundamental reality itself, existing objectively and independently of human thought. The aim of science would therefore be to discover and understand these mathematical structures. The success of scientific theories would be due to their (partial) correctness in capturing these fundamental structures. Especially in this interpretation of scientific realism—arguably inapplicable outside foundational physics—the no-miracle argument \cite{Putnam1981} develops its full persuasive power. From an instrumentalist perspective, however, mathematical equations and structures remain useful tools. These tools help make predictions and explain observable phenomena without being considered the actual reality. For instrumentalists, mathematical structures are constructs of the human mind that allow us to systematize and predict complex phenomena but are not ontological entities. Their utility lies in their practical applicability, not in any supposed objective existence.  \textit{The central assumption of the aforementioned model will be the combination of a) Worrall's structural realism and b) Tegmark's MUH, ultimately leading to a classical scientific realism with respect to mathematical structures (only) in fundamental physics}. We remark that it is decisive to begin with Worrall's weaker, epistemological position instead of directly acknowledging OSR, as Tegmark's MUH is in some sense restricted to fundamental physics: If the MUH is true, then there is no distinction between physical reality and mathematical structures in the most fundamental entities in the world, implying the collapse of ESR and OSR into a single view: the structure is all there is. However, van Fraassen (\cite{vanFraassen2006}, p.\,292) justifiably criticized such a move, if it would be projected on all natural sciences: "What has looked like the structure of something (...) is actually all there is to nature (...) and the difference between it and ordinary scientific realism disappears." Indeed, such a generally applied OSR seems inadequate and too simple in other disciplines of science—where physical entities have intrinsic properties beyond mere structure. A weakening à la Worrall, that only structural relations are \textit{knowable}, is a more appropriate realist view on science in general. Nonetheless, the aforementioned collapse, as we will outline now, appears to be present at least in fundamental physics, thereby heavily simplifying the plethora of (epistemic/ontic structural, effective, selective, pragmatic, semi-) "realisms" present in today's literature: for our purposes, we are in the fortunate situtation that van Fraassen's observation \cite{vanFraassen2006} is right and we end up with the classical dichotomy of scientific realism and instrumentalism as a simple and solid basis for our model.

Let us therefore bridge to QFT: Within the context of structural realism combined with the MUH, elementary particles can be seen as nodes in a network of mathematical relationships. These relationships arise from the structure of the underlying quantum fields and symmetries (fundamental interactions). The quantum numbers attributed to a particle (e.g., charge, spin, color charge) are not independent attributes but emergent properties resulting from the interaction within the mathematical structure\footnote{In QFT, interactions are not externally imposed processes but follow from the structure of the theory itself. For instance, the coupling of particles to fields arises from the requirement of gauge invariance of quantum fields. These interactions are not arbitrary but uniquely determined by the mathematical structure of the corresponding Lie groups.}. In this sense, the physical is not independent of mathematics but an emergent aspect of its structure. This supports the MUH in the sense that the mathematical structure not only \textit{describes} reality but is identical to it and is instead \textit{discovered} by physicists. The mathematical gauge symmetries of the Standard Model and their necessary and unique consequences for the shape of elementary particles are thus, by virtue of the MUH, one of the most compelling arguments for scientific realism. A consistent mathematical structure of elementary particles and their interactions (as intended to be formulated by AxQFT) would thus become the primary existing entity in the microcosm, not merely its descriptive tool. To put it provocatively, the electron is not a yellow sphere with a black minus sign on it, but appears to be only a set of rational quantum numbers: nothing more needs to be added to describe its ontological nature, and any physicality perceived in the mesocosm disappears in the microcosm, where pure mathematical structure is all there is. However, the remaining inconsistent structures of SM and BSM physics (as tolerated by CQFT) are equally compelling in showing that instrumentalism cannot be entirely displaced. Physicists may regard mathematical pragmatism as an efficient engineering tool, whereas philosophers may see unresolved inconsistencies as obstacles to truth-tracking. This paper suggests that these views need not be in opposition, and that the notion of a realism quotient offers a principled way to integrate both attitudes. With these two poles of QFT, we arrive at the tension between rigor and pragmatism creating frictions between realism and instrumentalism. 

\section{Rigor and Pragmatism in history of QFT}

It seems paradoxical: In BSM physics, significant concerns are addressed that, in the view of many philosophers of science and physicists, do not constitute inconsistencies in the theory of elementary particles but have nevertheless led to considerable efforts. These include the naturalness problems of the Higgs mass and the CP-violating $\theta$, as well as the grand unification of gauge interactions. This article, however, addresses inconsistencies that have raised no significant concerns: Physicists, accustomed to the success of the SM, have developed a remarkable ability to ignore this mathematical elephant in the room. The current debate about inconsistencies is, however, more vivid within the philosophy of science, to which this article aims to contribute. Dawid (2019) pointedly writes about the lack of realism or the instrumental nature of today's theoretical microphysics:
"Things are particularly difficult in theoretical physics where some theories are predictively highly successful even though they are known not to be true for conceptual reasons. QFTs are (...) difficult to appraise in terms of truth values."
Initially, in late 1940s, nothing seemed capable of halting the success of QFT after precisely describing the anomalous magnetic moment $g$ of the electron (\cite{Schwinger1948}, \cite{Dyson1949}, \cite{Feynman1950}). Continuous refinement provided, through the latest experiments and CQFT calculations, an even more impressive empirical adequacy of quantum electrodynamics, comparing recent experimental and theoretical values $g_{ex./th.}$: \\ \\
- $g_{\mathrm{ex.}} = 2,002 319 304 362 56 (35)$ $\qquad$ \cite{Fan2023} \\ - $g_{\mathrm{th.}} = 2,002 319 304 363 22 (46)$ $\qquad$ \cite{Aoyama2015}, \cite{Laporta2017} \\ \\
For this extraordinary agreement, there are scarcely any comparable success stories in science. It seems hardly worth mentioning that, in Worrall's sense, a multitude of mathematical structures must be inherited by a successor theory to the SM, and thus, from the no-miracle argument, a substantial portion of realistic structures in the SM can be inferred. Nevertheless, shortly after the development of quantum electrodynamics, indications of internal contradictions emerged that stood in the way of the triumph of realism: "Realists aim to explain a theory’s empirical success by appealing to its truth, but QFTs seem insufficiently rigorous, inconsistent, and ad hoc to be true" (J.\,Fraser 2020, p.\,392). Among these issues are the divergence of perturbative series \cite{Dyson1952}, the existence of a Landau pole \cite{Landau1954}, Haag's theorem \cite{Haag1955}, and the lack of axiomatizability of one of the fundamental building blocks of foundational physics (\cite{Wightman1956}, \cite{HaagKastler1964}), for example, with respect to the Poincaré covariance of operator-valued distributions.

While the singularities of perturbation theory only temporarily threatened the practical application of relativistic quantum mechanics to elementary particles, and renormalization subsequently paved the way for the unprecedented success of the Standard Model, other problems threaten the very existence of a strictly formulated relativistic quantum mechanics itself. Here, history repeats: At the end of the 19th century, the symmetries of classical electrodynamics (Lorentz covariance) proved incompatible with the transformation behavior of mechanics (Galilean covariance). The resolution of this incompatibility was special relativity. It remains to be seen what resolution the problem of the strictly speaking incompatible Poincaré covariance of quantum fields will bring in the future. Yet one thing becomes apparent once again, as D.\;Fraser \cite{Fraser2009} illustrates with other examples: "QFT is hardly unusual in requiring refinements over a period of time. There are many cases in which an inconsistent formulation of a theory was replaced by a consistent one in the course of its historical development."
\\ \\
After noting these severe issues, an overwhelming majority of physicists subsequently turned their attention to the practical application of QFT to the evolving SM and, until today, theories beyond it—thus practicing CQFT. A small group took on the mathematical difficulties of QFT and established a mathematically rigorous version: AxQFT. To this day, it is not expected that the work of one research program could contribute to solving the problems of the other, as AxQFT has yet to produce a physically relevant model in four dimensions: "Rather than seen as rivals, (they) are complementary projects within the broadly construed research program of QFT" (\cite{Athanasiou2022}, p.\; 61). To later defend our "normative prospection", we must first clarify why the past successes of AxQFT nevertheless justify ascribing to it a closer relationship with scientific realism.
In general philosophy of physics, this is evident in the tendency to use AxQFT for ontological work on elementary particles \cite{Kuhlmann2010}, even though it has yet to produce empirically adequate models. In philosophy of science, this is reflected in the almost self-evident assertion that the world itself cannot be inconsistent, and thus the inconsistent CQFT still contains instrumental mathematical structures that AxQFT seeks to replace with realistic structures. Let us delve into some details.
\\ \\AxQFT can, strictly speaking, be divided into two research programs (based on Wightman, and Haag-Kastler, respectively), which attempt to rigorously formulate these properties using different mathematical languages and emerged in the 1950s. Both address the question of how a world possessing both quantum mechanical and special relativistic properties simultaneously can be described. The outcome of these efforts remains abstract axiomatizations into which the successful, empirically confirmed theories of the Standard Model cannot be integrated. In both programs, the identification of basic entities and a set of axioms acting on them was achieved as early as the formative days of QFT, during which QED was also formulated as a physical QFT. However, proof is still lacking that physical theories like QED or even more complex, non-abelian theories of the Standard Model can be reformulated in the axiomatic languages\footnote{The Wightman approach to QFT shows particular strength in constructing models. Based on Wightman and Osterwalder-Schrader, several lower-dimensional QFT models adhering to strict axioms (\textit{toy models}) have been formulated. By contrast, the algebraic approach of Haag-Kastler draws on profound theorems from the general theory of operator algebras and can even be applied to the curved spacetime of general relativity. Generality and flexibility are the terms that best characterize this research program: it enables the rigorous proof of fundamental mathematical relationships in the universe. As Haag \cite{Haag1996} put it, "the algebraic approach (...) has given us a frame and a language not a theory." These include the CPT theorem (mentioned in the introduction, recalling the invariance under simultaneous time reversal, spatial reflection, and charge conjugation) and the spin-statistics theorem. The program is also of interest to philosophers with an ontological focus, as it allows for the proof of certain no-go theorems that have implications for the particle concept in QFT (\cite{ReehSchlieder1961}, \cite{Malament1996} ). A recent introduction to the research program can be found in \cite{Halvorson2007} and \cite{FewsterRejzner2019}.}.  \\ \\ Philosophical considerations regarding the current state of QFT are connected to the broader realism-instrumentalism debate, initially embodied in the "Fraser-Wallace debate": D.\,Fraser argued that fundamental QFT should strive for mathematical consistency, even if this means initially sacrificing empirical adequacy for conceptual clarity. Wallace, on the other hand, argued that QFT in its modern application should be seen as an effective theory, where empirical adequacy is more important than the claim of fundamental validity. These opposing views shaped the discussion for over a decade. However, the debate has increasingly yielded compromises. In her summary of the discussion, Athanasiou (2022) paints a nuanced picture of the debate, showing that a gradual rapprochement is emerging in the philosophical consideration of QFT\footnote{Important contributions include those by Cao \cite{Cao2003}, Earman and D.\, Fraser \cite{EarmanFraser2006}, D.\,Fraser (\cite{Fraser2008}, \cite{Fraser2009}, \cite{Fraser2011}), Wallace (\cite{Wallace2006}, \cite{Wallace2010}), Kuhlmann \cite{Kuhlmann2010}, \cite{Baker2014}, Swanson \cite{Swanson2017}, Miller \cite{Miller2018}, Ruetsche \cite{Ruetsche2018}, \cite{Williams2019}, Dawid \cite{Dawid2019}, J.\,Fraser (\cite{FraserThesis2016}, \cite{Fraser2020}), Rivat \cite{Rivat2021}, and Dougherty \cite{Dougherty2023}. We apologize to all whose contributions have been overlooked.}. This rapprochement between CQFT and AxQFT is to be further advanced in this article. 
\section{How Much Rigour Do We Need for Empirical Progress?}
\subsection{Realism Quotients}
For a convincing compromise between scientific realism and instrumentalism, which can offer a way out of the dilemma of the two diverging programs of QFT, we begin with the rational reconstruction of modern\footnote{Even in the early days of Leibniz and Newton, there was a vivid dispute about the reality of mathematical structures utilized for classical mechanics such as infinitesimal intervals in time and distance. This debate is somewhat disconnected from the mathematical issues of modern foundational physics. We restrict ourselves to the developments of the 20th century in order to not make the model overly complicated.} foundational physics from the introduction (\textit{descriptive retrospection}). Tailored to mathematical realism (MUH), the following insight is  central: A mathematically inconsistent and therefore "overcomplete" rather than incomplete theory may contain substructures or statements that correspond to reality, even if the overall picture is flawed. By combining Worrall's structural realism with Tegmark's MUH recapitulated in Chapter 2, it is possible to label selected mathematical constructions as real.
Reacting on Laudan's critique by means of the pessimistic meta-induction (PMI), one can concede that from the currently best theories, which according to PMI are not actual representations of reality, elements can be extracted that are likely to survive the evolution of theories and will be central components of future improved theories. They are, in a sense, the fixed points in the dynamic system of changing theories. Given the obvious limitations of the Standard Model (SM) to certain energy scales (effective field theory), a pure scientific realism that sees the entire model as (at least approximately) true is, of course, unjustified. This does not even require the PMI argument—it is a provisional model by construction that will one day be discarded (cf.\;\cite{Hartmann2001}).
Yet whether the future replacement of the theory as a whole is known or not: One can escape Laudan's pessimism and argue that realistically interpretable aspects of the theory were the reason for its empirical adequacy. These aspects are attributed lasting value for theory formation and are also seen as guarantees of success for future theories. The relevance of these aspects to the overall picture is undoubtedly significant despite the provisional nature of the theory. Dougherty (\cite{Dougherty2023}, p.\;1) states: "The dramatic successes of QFTs like the Standard Model of particle physics must then mean that their descriptions of the unobservable are substantially correct. We should be realist about them if we’re realist about anything." Indeed, the no-miracle argument unfolds its full plausibility with the ever increasing experimental resolution: an accidental congruence of the theory with reality becomes ever more unlikely when building larger colliders and finer detectors of unprecedented precision. Restricted to the consistent, unquestionable mathematical descriptions (structures) by CQFT, we align with Dougherty. The remainder, however, needs to be labelled instrumentally. In QFT, many mathematical structures bear clear signatures of being merely instrumental. This stands in contrast to more ambiguous realist/instrumentalist distinctions in other sciences (e.g., biology or psychology), where the structures are less sharply formalized and empirical precision is limited. In QFT, however, a distinctive melange of epistemic symptoms—adhocness, pragmatic compromises, mathematical inconsistency, and subsequent historical disposability—make the instrumentalist character of some structures unusually transparent. Thus, instrumentalist structures are not just assumed—they are often exposed. This clarity allows the estimate of the share of instrumental elements in different QFT models to be more precise, historically traceable, and empirically anchored than in other domains. These signs allow for a clearer mapping of such elements onto the diagram proposed in Fig.\;1 that we will explain now.

Extending this thought (independent of the SM), to any given point in scientific understanding, we propose a ratio or quotient $R$ between these realistic and instrumentalist mathematical structures in the foundational physical theories, which seems to increase with the finer resolution of reality in experiments and the associated evolution of theories:
\begin{small}
\begin{align*}
R := \frac{\mathrm{number\,of\,realistic\,mathematical\,structures}}{\mathrm{number\,of\,all\,(instr. + real.)\, mathematical\,structures}}
\end{align*}
\end{small}
 A schematic and simplified visualization\footnote{Of course, monotony and linear increase of $R$ do not match with the nonlinear history of physics with its loops and dead ends. For simplicity, we refrain from details and concentrate on the clarity of the recommendation for future theory development. Numerical values of $R$ can hardly be specified in principle—it serves rather as a heuristic vehicle to model the theory development of QFT (and, more generally, foundational physics) than as a precise, determinable value. The limits of a too mathematical interpretation of $R$ are obvious: A more complex theory of the future showing a better empirical adequacy may contain both more instrumental and realistic element, and even the unique countability of these elements may be doubted.} is given in Fig.\;1: Each improved theory of foundational physics contains more realist elements than its predecessor, which survive theoretical shifts—they successively replace pragmatically motivated, instrumentalist elements. Mathematical pragmatism—used at the expense of mathematical rigor but for easier handling of the theory—prevents empirical adequacy in a mathematical universe after a certain point: too many instrumental structures are tolerated. In provisional effective theories, instrumental sacrifices must always be made by definition. Yet, when balanced correctly, not at the expense of empirical adequacy (this balance shall be defined as $R_{ea}$). 

To prevent terminological misunderstandings: \textit{This conception generally replaces the antithetical relationship between realism and empirical adequacy in the sense of van Fraassen} by understanding the latter as a dynamic property of "fitting the data," which is continuously challenged in each era by new experimental precision and can at times provoke theory change. Empirical adequacy persists the longer, the greater the theory’s closeness to truth (verisimilitude), expressed through $R$ as a dynamic formulation of the no-miracle argument: While van Fraassen was to regard this property as static, a permanent label assigned to a theory—thereby rendering any discussion of realism and truth superfluous— in the model discussed here, the gradual end of the empirical adequacy of established theories can be explained by the continuous approach toward truth enabled by new technological advancements. Such an approach is particularly illustrative in particle physics, as it represents a continuous zoom into the microcosm revealing, order of magnitude by order of magnitude, a progressively more resolved view of its true nature.

In the general debate, Laudan \cite{Laudan1981} also criticized the assumption that scientific theories over time approach an ever more accurate picture of reality, as the assumption of a growing realism quotient suggests. He points to cases where successful theories were replaced by completely new ones that showed no continuity with the mathematical structures of their predecessors (incommensurability in theory dynamics), which contradicts the idea of convergent and cumulative truth. However, the incommensurability of current theories to predecessor theories is nothing but an increase in the realism quotient $R$ in the sense that realist elements are, by definition, incommensurable with some previously utilized instrumental elements that are not compatible with the mathematical structures of reality but were compatible with the limited degree of experimental precision.
This notion, the concept of this article, does not endanger the convergence concept but simultaneously escapes Laudan’s criticism. The convergence concept must merely be restricted to the correct interpretation of the limit $R \to 1$: a steady approach to truth only means the \textit{final abandonment} of more and more instrumental elements while maintaining realist elements—not the \textit{continuous refinement} of purely instrumental theories striving towards truth.
\begin{figure}[h!] \centering \includegraphics[width= 1.0\textwidth]{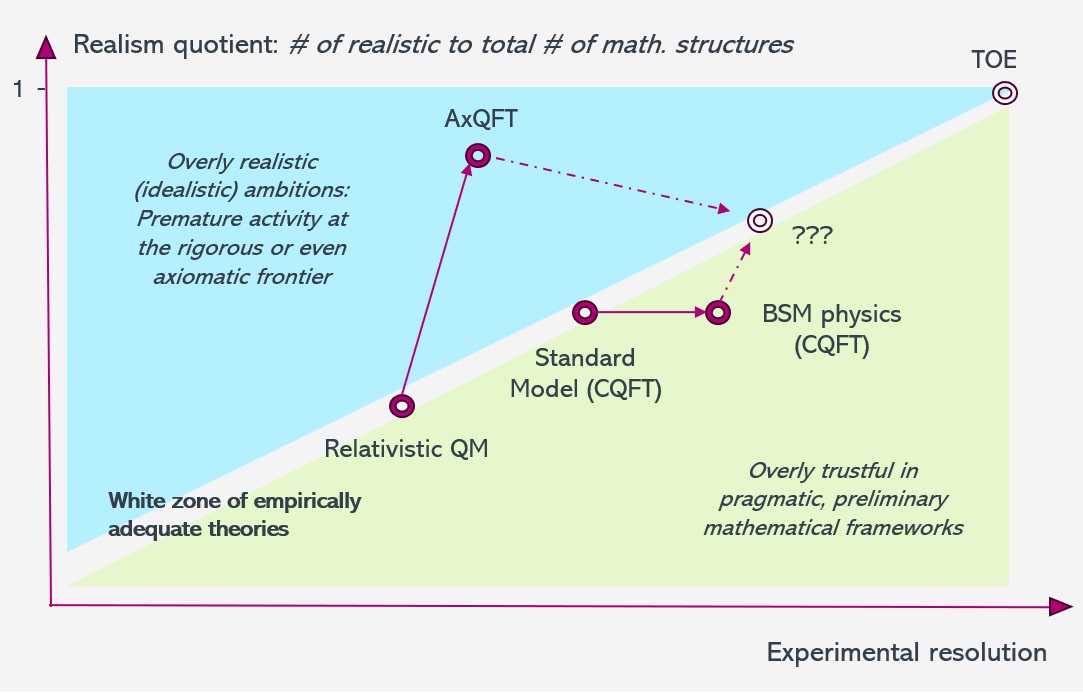} \caption{Reconstructive, simplified scheme of progress in quantum field theory so far, and normative guidance for future progress. \\
\textit{x-axis:} describes the increasing resolution/precision of experiments (thus also implicitly representing the time arrow of theory development). \\
\textit{y-axis:} represents the (presumed) proportion of realist elements in each theory. \\ 
\textit{White zone:} In all epochs of scientific progress, there is a required balance between realism and instrumentalism for empirical adequacy ($R\approx R_{ea}$), which is recalibrated at all times based on the experimental level of development. This realism quotient increases throughout the history of foundational physics and could culminate in a TOE at a value of 1 (upper right corner). \\
Deviations upward or downward from the white zone indicate overly idealistic demands on the mathematical structures (hoping they correspond to real structures of the world) or too high demands on instrumentalism (overestimating the long-term success of mathematical pragmatism). \textit{Figure created with Microsoft Powerpoint 2023}
 \label{fig1}} \end{figure}
Fig.\;1 also shows this limit: the hypothetical end of the growing realism quotient. In this case, humanity would have reached a "Theory of Everything" (TOE) with $R=1$, a convergence in finite time scales, if this limit exists. Crowther \cite{Crowther2019} formulated some properties of such a fundamental theory. These properties include, among others, an expression of the fact that the pragmatic concessions of CQFT (instrumental component of the realism quotient) disappear: A TOE is non-perturbative (exactly solvable), unified, singular, internally consistent (well-defined formally, with no problematic singularities), and level-comprehensive (no gaps and no overlap)\footnote{A TOE remains a "TOE" as long as no deviations from these properties are noticed at a later stage of research. Crowther’s formulation thus always offers only provisional certainty that one might have actually found such a TOE.}. With Crowther’s criteria, it is again clarified why the consistency efforts of AxQFT may lead to a higher realism quotient, compared to the tolerance of inconsistencies and rigor issues in CQFT. The latter shows a perturbative structure with limited potential for further unification, problematic singularities (Landau) and gaps to even special relativity when aiming for an axiomatic construction (Poincaré covariance). AxQFT, on the other hand, provides exactly solvable models in lower dimensions, shows internal consistency and is constructed to be level-comprehensive, allowing perhaps even to close some gaps to the structures of general relativity. AxQFT seems to march towards structures of a TOE and to come closer to $R=1$, at least in comparison to CQFT. Of course, even with the completion of BSM physics including a more rigorous QFT, it would not be a TOE—in this research program, only the penultimate\footnote{Final questions concern quantum gravity, which is its own research program and requires completely different mathematics. BSM physics restricts to everything that can be constructed with QFT only, not requiring general relativity.} questions about the world can be addressed—at extremely high resolution of the microcosm. This undertaking alone is ambitious enough and claims to uncover some of the fundamental mathematical structures of the world. But surprisingly, the BSM physics research program adopts the same mathematical pragmatism as CQFT ($R_{SM}=R_{BSM}$), which was successful in the SM but has so far failed with empirical adequacy in Grand Unified Theories (GUTs), supersymmetry, or dozens of other BSM theories. This pragmatism is remarkable, if not naive: there is a comparable number of orders of magnitude between the Higgs boson and classical mechanics (regarding energies and length scales) and between the GUT scale and the Higgs boson. Mathematical structures in Newtonian Mechanics and CQFT describing the Higgs mechanism could not be more different. It is thus a highly bold conjecture that CQFT is still valid at the GUT scale, a dozen orders of magnitude beyond the reach of our largest accelerators.

The underlying, though mostly unspoken philosophical assumption behind this conjecture is that instrumentalism itself is unproblematic for the explanation of such fundamental physics. Mathematics is used as a useful language but not discovered as a real entity in the microcosm. Is this assumption, this pragmatism, justified in light of the fundamental nature of many BSM physics questions and the nature of modern particle physics as explained in Chapter 3? Or does the lack of mathematical rigor hinder scientific progress beyond the Standard Model\footnote{A side note may be helpful: Just because MUH may be correct, the exclusion of any instrumental mathematics for a pre-final theory is not necessary (only for a final "Theory of Everything", if it exists) for it to be empirically adequate. Otherwise, physics would never have been able to celebrate its successes in classical physics, nor still in the Standard Model (PMI).}?

The concept of the increasing realism quotient after finishing the SM research program suggests a nuanced answer: Fig.\;1 shows how the continued use of the previously sufficient CQFT in BSM physics does not increase the realism quotient and thus causes the successor program to the SM to fall outside the desirable zone of empirical adequacy ($R_{BSM}<R_{ea}$). This representation is, of course, to be understood purely illustratively: While the realism quotient is not merely a reflection of the increasing precision of theories, it is a tool for examining the balance between mathematical rigor (realism) and pragmatic concessions (instrumentalism) at all times in theory development.  

Let us now contrast CQFT with AxQFT, recalling the respective strengths and weaknesses of the research programs from Chapter 3. In this light, CQFT appears as a theoretical program that operated well within the white zone of empirical adequacy during the investigation of the Standard Model but may have lost touch with more fundamental principles in specific contexts of BSM physics, thus becoming outdated. D.\;Fraser (2009, p.\;20) thereby notes: "Resting content with (CQFT) because it is empirically adequate at large distance scales would be a strategic mistake because it would hinder the search for theory X." In contrast, AxQFT, due to its lack of empirical applications (no formulation in four dimensions), currently "hovers" above the white zone and might be thus conceptually and mathematically ahead of its time. 

Once again: the assumption that AxQFT has a higher realism quotient is based on the hypothesis that mathematical consistency is in any case closer to truth than pragmatic, but inconsistent models like CQFT. At first glance, this assumption appears paradoxical when Rivat (2020) writes about unrealistic theories: "In particular, some philosophers highly concerned with the infamous mathematical issues plaguing realistic QFTs have been drawn to believe that mathematically rigorous yet highly unrealistic QFT models in lower dimensions give us more reliable ontological information about the world than the heuristic yet empirically adequate QFT models in high-energy physics." However, such misunderstandings are likely of a primarily linguistic nature: a theory from AxQFT is labeled unrealistic and thus useless for the ontology of the microcosm because it is (as of yet) not empirically adequate. This attribution only considers the short-term success in past experiments and the short-term failure of AxQFT in constructing four-dimensional models. It overlooks, however, the long-term potential of AxQFT to project those mathematical structures—whose consistency makes them so attractive for ontology—onto the empirically adequate models of high-energy physics, without compromising their empirical adequacy. The fundamental mathematical building blocks and thus the ontological foundation of future QFTs, so the well-founded hope of AxQFT, will always remain the same—whether in a simplified \textit{toy model} or a practical model of high-energy physics. Therefore, the ontological potential of AxQFT, as discussed in \cite{Rivat2021}, is being considered on the wrong level, one that focuses only on short-term success. However, it is clear that the slow progress in AxQFT calls for patience, and its realistic interpretation may only fully unfold its plausibility in the more distant future—an outlook we wish to endorse (\cite{Kuhlmann2010}, p. 1634)\footnote{The realistic interpretation of AxQFT already goes back, albeit in a less elaborated form, to \cite{Saunders1988}, which has been regularly reproduced and developed further in the philosophical debate.}: "On a realistic reading, AQFT says something about the ultimate building blocks and the structure of the world. (...). One expects the formalism to somehow represent how nature works." Reliable ontological information, according to this view, cannot be provided by today’s high-energy physics (ibid.): "The problem with standard QFT is that it mixes different, mutually incompatible ontologies in an intransparent way." This hypothesis is primarily supported in this work by two arguments: first, by the ontological priority of consistent mathematical structures, which AxQFT aims to formulate, and second, by the long-term significance of stricter theories in the history of foundational physics. While the first argument has already been explained in detail above, the second will be substantiated in the next subsection. This (lack of) suitability of the two programs for ontological claims is ultimately the deeper reason for their conceptual distance, which has led to their respective isolation.

The complementarity of these two approaches to QFT also illustrates how theories that lie below or above the white zone can complement each other: While CQFT drove the empirical success of the last century, the AxQFT preserved fundamental insights that could one day find their place in a future theory of physics. In a sense, the AxQFT represents the long-term, as yet unrealized potential of scientific theories. With regard to AxQFT, one should indeed speak of premature activity on the axiomatic front, whose value will only prove itself in the distant future of fundamental physics. Complementarily, the CQFT embodies the short-term success of theories but now seems endangered in its universal role as the mathematical framework of particle physics due to the failures of BSM physics.

This brings us to the \textit{normative prospection} outlined in the introductory chapter: Ultimately, both programs fail because of their overly radical interpretations of realism or instrumentalism. A middle ground, ergo an appropriate realism quotient $R_{ea}$, seems desirable as a way out of this crisis—as schematically indicated by arrows in Fig.\;1. The challenge for future researchers, therefore, lies in finding the right balance in QFT between pragmatic applicability for concrete, contemporary particle physics and long-term conceptual rigor that paves the way for successor theories. Activating the dialogue between the two research programs is a central aim of Chapter 4.3.
However, let us first deepen the concept from Fig.\;1 with some more general reflections, which can always be illustrated and exemplified with CQFT and AxQFT.

In an ideal world, good theories would be designed so that they are always near or within the white zone of perfect empirical adequacy, i.e., the current experimental feasibility, at the time of their formulation in the history of science. However, even larger deviations from this ideal state, far above or below this white zone, can produce long-term successes in fundamental physics. In summary, three categories arise, which we want to explain below: 
\begin{itemize} 
\item[i)] a "just-in-time theory" with $R\approx R_{ea}$, a balanced ratio of realistic and instrumental structures at the time of development, measured by the experimental state of the art. 
\item[ii)] a "delayed workaround"  with $R<R_{ea}$, pragmatically motivated with too many concessions to be empirically adequate at all accessible scales at the given time (effective theories on purpose, limited to some scales where the model shall "effectively" work).
\item[iii)] a "premature wundertuete" with $R>R_{ea}$, idealistically motivated. They are subsequently recognized as either (a) visionary and realistic, thus being "hyperrealistic" in the time they were formulated, or (b) physically irrelevant and dismissed as mathematical gambling. \end{itemize}
If a theory in the last category prematurely includes elements that can justifiably be assumed, but not confirmed to be real parts of our world, this necessarily happens in a state of epistemic agnosticism and thus outside the white zone. Such openness at least offers the chance that the theory will later be recognized as correct and realistic if its structure proves itself in further empirical and theoretical development. The situation is entirely different when a theory of category ii) lingers in excessive pragmatism for too long. In this case, there are no connection points to the empiricism that has meanwhile advanced, thereby paralyzing the entire theoretical program. Such a theory remains overly instrumental for all time and ultimately becomes irrelevant because it does not allow progress and does not offer the possibility of identifying or integrating new realistic elements.
Thus, while a certain optimism in the early inclusion of potentially realistic elements can be risky but fruitful, excessive pragmatism is a dead end in the long run according to our mdoel. History of science shows that some theories that initially appeared empirically agnostic later proved their realistic elements (see Chapter 4.2). Others, however, were recognized as errors. 

Both theories of categories ii) and iii), far below or above the white zone of empirical adequacy, are anachronisms in their own way. Thereby, it becomes clear that the success or failure of a theory depends not only on its internal consistency and empirical adequacy but also on its temporal appearance in the history of science: Theories that are situated below the white zone seem regressive because they adhere too strongly to mathematical structures that, measured with today’s experimental precision, appear outdated. Such theories offer little room for progress and sooner or later contradict new empirical findings. Theories located above the white zone, conversely, are either groundbreaking some day or just leaps into the void, but due to a temporal misplacement, they cannot be transferred into empirical research: They are ahead of their time and can only unfold their full significance long after their formulation—or be dismissed as meaningless or physically irrelevant.

Inspired by \cite{Fraser2009}, we would prefer not to refer to these theories as \textit{idealizations}, since there is a risk of confusion between idealization and pragmatic simplification in the sense of instrumentalism, but rather as \textit{idealistic}. Those who seek to formulate such theories are evidently striving for the ideal of perfection—a contradiction-free, ultimately valid, and thus ontologically useful description of the world. Idealistic theories are based on the assumption that there is an underlying, often mathematically elegant structure that describes reality. These theories place strong trust in the power of ideas, concepts, and mathematical structures, even if their empirical confirmation is still pending. Such theories often postulate entities whose realization in our universe remains unclear. Their evaluation depends on the eventual empirical confirmation of these entities. They share with realism the conviction that there is an objective structure to be discovered by the theory, rather than merely described. As already noted, it is extremely difficult ex-ante to assess whether an idealistic theory actually contains more realistic elements than its predecessor and the presence of realistic elements at all.  

The upper zone of such hyperrealistic ambitions in Fig.\;1 must therefore be considered a danger zone, with the risk that theories here may never become empirically accessible or turn out to be physically irrelevant. In this case, the term of an excessively, thus too high \textit{realism} quotient is misleading, as the postulated structures have never revealed themselves as \textit{real} by current experiments. There is simply the chance of being confirmed to be real one day by future experiments. An overly high realism quotient in such theories creates a surplus structure of information, which is at risk of being filtered out by the lex parsimoniae (Ockham's principle) without jeopardizing empirical adequacy. According to this parsimony, they are unnecessary or even counterproductive at the given time because they unnecessarily complicate the theory that should not postulate more entities or structures than absolutely necessary. At first glance, one might therefore argue that theories with surplus structures should be discarded, compared to their simpler alternatives. If a theory introduces unnecessary complexity without offering additional empirical gain, it contradicts the demand for simplicity.

Historically, there are several cases where theories with initial surplus structures later became crucial for scientific progress. This should serve as an explicit warning in dealing with the problems of QFT. It may only become clear decades into the future that these structures provide essential insights and are necessary for the further development of physics. It is a fortunate fact that scientists are sometimes willing to "invest" in a theory that has (potentially useless) surplus structures, even if these are not currently empirically verifiable. Such investments may later yield enormous scientific returns and thereby relativize Ockham's Razor. We delve a bit deeper into a cautious filtering of these surplus structures in Chapter 4.3, but first look back into the insightful history of particle physics.

\subsection{Historical Justification} 

In the development of theories, there has regularly been an overreach in both directions—above and below the zone of theories with empirically adequate realism quotients—initially and directly at the expense of empirical adequacy. A selection of four examples, categorized in the graphic from Fig.\;2, is intended to legitimize the application of the model for normative prospections in BSM physics in particular and in the philosophy of science in general:

\begin{itemize} 
\item Fermi's theory of weak interactions as an example of an inconsistent (instrumentalist) mathematical structure, which ultimately compromised empirical adequacy: Category ii) 
\item Gell-Mann’s Eightfold Way (early quark model: a hyperrealistic theory of nucleons in these days appearing to be an artifactual structure without reference to reality): Category iii), path a) 
\item Heisenberg’s World Formula as an example of a mathematically elegant but physically irrelevant theory, never empirically adequate: Category iii), path b)
 \item Connes' noncommutative SM as a prospective example with a presently non-testable spacetime structure (hyperrealistic claims) but with an empirically adequate prediction of the Higgs mass: Category iii), physical significance still unknown, path a) or b) \end{itemize}

\begin{figure}[h!] \centering \includegraphics[width= 1.0\textwidth]{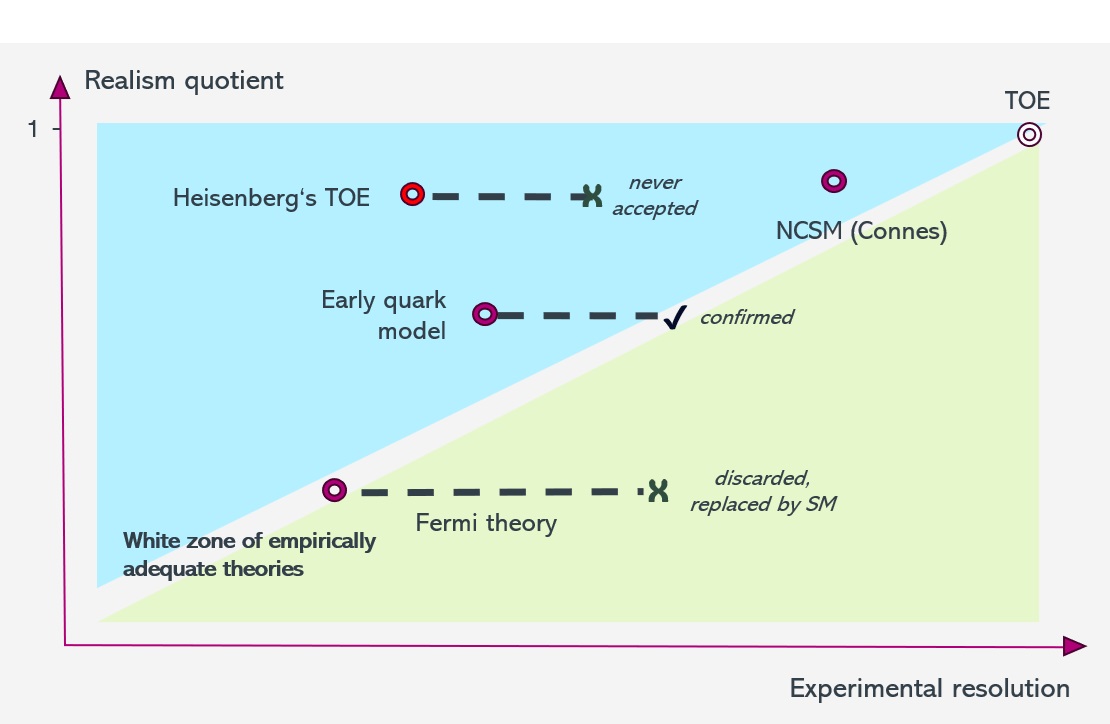} \caption{Historical examples from particle physics illustrate the necessity of a balanced relationship between realistic and instrumental mathematical structures in fundamental physics, which must be regularly adjusted depending on experimental precision. \textit{Figure created with Microsoft Powerpoint 2023} \label{fig2}} \end{figure}
\textbf{Fermi’s Theory of Weak Interactions:} Fermi’s theory of weak interactions was an initial attempt to describe the weak nuclear force through an effective theory. In this model, the interaction is represented by a four-fermion coupling \cite{Fermi1933}. It proved extremely useful in explaining phenomena such as beta decay, but it was not renormalizable, meaning it produced inconsistent (infinite) predictions when calculations went beyond the leading order. Fermi’s theory is a prime example of a theory with a low realism quotient\footnote{The first version of Fermi's publication was rejected by Nature on the grounds that it "contained speculations too remote from reality" \cite{Close2012}.}: It was clear that there cannot be a direct interaction of four fermions in our universe, making the Fermi interaction not fully empirically adequate. However, this makeshift theory was quite successful as it correctly described many observations of weak interactions at the then-current experimental resolution of the microcosm (placing the theory in Fig.\;2 very close to the white zone with its specific $R_{ea}$ of Fermi's time). However, it did not provide a deeper mathematical structure revealing the true nature of the weak interaction and was eventually superseded by the electroweak theory in the SM, which introduced W and Z bosons as mediators of the weak interaction. The theory was empirically adequate within a limited energy range but could not be extrapolated to higher energies, where the actual existence of the mathematical structure of the electroweak model manifested as measurable W and Z bosons. It lost its validity. Thus, Fermi’s theory must be classified in Category ii)—the long-term fate of all theories with too many instrumental components.
\\ \\
\textbf{The Eightfold Way:} In the early 1960s, Murray Gell-Mann introduced the concept of the Eightfold Way, an $SU(3)$ symmetry, to explain the apparent order in the "zoo-like" diversity of hadrons then known (\cite{GellMann1961}, \cite{GellMann1964}). The emergent quarks, as principally measurable (as recognized later on) representations of this mathematical structure, were postulated as hypothetical building blocks of hadrons, endowed with properties such as charges equal to one-third or two-thirds of the elementary charge; a concept that seemed revolutionary and highly speculative at the time. Gell-Mann himself described the quarks in early publications as a mathematical framework intended to enable an elegant grouping of particles without necessarily claiming physical reality. Quarks contradicted the prevailing image at the time, which described protons and neutrons as point-like (thus creating a surplus structure of an overly complicated microcosm). Their properties, such as fractional charges, appeared inconsistent with existing experimental physics, and the scientific community reacted with skepticism toward these unobservable entities. Simultaneously, Gell-Mann’s theory was recognized as a significant advancement due to its ability to bring order to the chaotic diversity of hadrons. Nevertheless, the high idealistic claims about the structure of protons were still above the zone of empirical adequacy, converse to Fermi’s theory: some time had to pass before the realism quotient was rebalanced. In the 1970s, deep inelastic scattering experiments led to the discovery that protons and neutrons indeed consist of smaller point-like objects corresponding to Gell-Mann’s quarks. The Eightfold Way must therefore be retrospectively recognized as a "treasure chest" ahead of its time (Category iii), path a). \\\\
 \textbf{Heisenberg’s World Formula:} In the 1950s, Werner Heisenberg developed the idea of a TOE that would describe all elementary particles and their interactions through a unified mathematical structure (\cite{Heisenberg1967}, \cite{Blum2019}). Heisenberg’s approach is an example of an overly high "realism quotient" that was empirically unfruitful and likely never will be. Heisenberg’s theory aimed to reveal a deeper truth about nature, but its highly idealistically motivated mathematical and conceptual requirements rendered it empirically inadequate in light of subsequent developments in particle physics. His ambition assumed that a single mathematical structure could fully describe all elementary particles and interactions. However, the real nature of the world proved to be more complex and less accessible to such a universal mathematical description. The World Formula ultimately failed because it was too focused on an idealistic claim, without providing a viable instrumental framework for empirically testable predictions. Theories with an excessively high "realism quotient" occasionally fail because they are based on an overly idealistic approach to foundational physics or an overly speculative notion of the real nature of the world at the time of their development, as outlined in case b) of Category iii). Such theories tend to place very high demands on mathematical rigor and consistency, attempting to capture deep, fundamental truths about nature. However, this ambition can lead to theories that are too far removed from empirical data or experimental verifiability, ultimately leading to their failure and rendering the wording of a \textit{realism} quotient absurd\footnote{The Kaluza-Klein theory for unifying classical electrodynamics and general relativity in a five-dimensional manifold \cite{Kaluza1921} can serve as an example outside particle physics.}. However, it can also turn out differently, as the final example is meant to illustrate. \\ \\ 
\textbf{The Noncommutative Standard Model:} Connes’ approach of noncommutative geometry represents a fundamental extension of the SM of particle physics by describing spacetime as a noncommutative space. In this theory, classical geometry, as used in General Relativity, is replaced by a mathematical framework based on operator algebras. This concept allows spacetime itself to be viewed as an algebraic structure in which coordinates no longer commute. The key idea is that such a noncommutative spacetime could provide a natural platform for unifying quantum mechanics and gravity. This is a profound ontological assumption that extends beyond the established empirical basis—unobserved so far, introduced as an ideally motivated mathematical structure. Despite the speculative nature of this approach, Connes’ theory has proven to be remarkably empirically adequate. One of the greatest achievements of noncommutative geometry was the prediction of the Higgs boson mass. Connes demonstrated the consistency of his model with a Higgs mass of approximately 125 GeV (\cite{Connes2010}, \cite{Connes2012})\footnote{Initially, the model predicted a Higgs mass of 160–180 GeV by neglecting a real scalar field coupling to the Higgs field \cite{Connes2010}. Incorporating this field corrected the value downward, later consistent with the experimental value. This ex-post success was largely ignored by the scientific community.}. Although it is still considered an idealistically motivated theory today, Connes’ approach shows that high conceptual ambitions do not necessarily preclude empirical adequacy in other respects. However, the assignment to consequences a) or b) of a theory from Category iii) is not yet possible.

 \subsection{Outlook: Dissolving Intellectual Isolation}

The current crisis in BSM physics may be reinterpreted in light of the discussed model, potentially adding another cause to the crisis and also giving a solution to a phenomenon we want to call the \textit{intellectual isolation of QFT research programmes}. The lack of experimental breakthroughs—the absence of empirical confirmations for all existing theories—raises the question of whether this stagnation is not only due to technical or experimental limitations but also to an inappropriately pragmatic approach to the mathematical foundations of BSM physics. The focus on sometimes arbitrary problem-solving within the mathematical framework of the predecessor theory—the Standard Model (SM)—(e.g., the ad-hoc introduction of additional symmetries) could thus be a short-term but ultimately insufficient strategy. For instance, Hossenfelder \cite{Hossenfelder2020} advocates for a new approach that relies more on mathematical coherence and conceptual depth, too, not only restricting her statement to BSM physics. Previous theoretical breakthroughs "have in common that they were based on theoretical advances which resolved an inconsistency in the then existing theories. (...) Some of the inconsistencies in the current theories are the missing quantization of gravity, the measurement problem in quantum mechanics, some aspects of dark energy and dark matter, and some issues with QFTs."

If BSM theories primarily aim to solve existing problems of the SM instrumentally while tolerating all known inconsistencies of CQFT, they risk remaining stuck in a mathematically inadequate framework for the depth of the questions. This "business-as-usual scenario" may have seemed fruitful in the short term due to the outstanding adequacy of the SM but may be blind to deeper structural insights. Just as the 4-fermion coupling in the theory of weak interaction was blind to the gauge symmetry of the electroweak interaction, today's provisional constructions that served the SM could lead BSM physics into a dead end.
\\ \\ 
To achieve empirical adequacy in future BSM physics theories, we would like to add another suggestion to the numerous already proposed. This \textit{normative prospection}, already mentioned in the introduction, was schematically outlined in Fig.\;1. Let us now turn to some details: According to the insights gained from the rational reconstruction of modern foundational physics, both CQFT and AxQFT may abandon their extreme positions in favor of a more balanced approach (of course, a successful BSM theory of some future might consist of completely different mathematical structures than those of CQFT and AxQFT. There is always an alternative to the arrows drawn in Fig.\;1). The conventional approach should place more value on mathematical consistency, while the axiomatic approach needs to become more flexible and experimentally oriented. A synthesis of both programs could be the key to developing a new generation of BSM theories that provide deeper insights into nature and are empirically adequate. Heisenberg (\cite{Heisenberg1967} , author's translation, p.\;V) himself assists us in this regard, succinctly summarizing the \textit{normative prospection} from our model, albeit for his TOE, which failed for the aforementioned reasons: "At the current stage of the theory, it would be premature to begin with a system of well-defined axioms and derive the theory from them using exact mathematical methods. What is needed is a mathematical description that is appropriate to the experimental situation that does not seem to contain contradictions, and that can therefore perhaps later be developed into an exact mathematical scheme."

The risk of failure in our current recommendation becomes evident in a later statement by Heisenberg, where he concludes: "The abundance of experimental material on elementary particles (...) may be considered a sufficient reason to assume (...) that all the important facts needed for formulating the theory are indeed known" (ibid., p.\,134, authors translation). Nevertheless, proponents of AxQFT express views similar to our conclusion based on the model in Fig.\;1: "The main problem of QFT turned out to be to kill it or cure it: either to show that the idealizations involved in the fundamental notions of the theory are incompatible in some physical sense, or to recast the theory in such a form that it provides a practical language for the description of elementary particle dynamics" (\cite{StreaterWightman2000}, p.\;1). Effective realism \cite{Williams2019} is also very helpful for bringing AxQFT closer to the white zone of empirical adequacy. It provides concrete guidance to identify and remove artifacts in AxQFT\footnote{He argues that the problematic aspects of CQFT can be confined to extremely high energies, which lie outside the theory's scope. Indeed, this aptly describes the extent of permissible pragmatic concessions. In the absence of empirical access to this energy range, these aspects can be deferred to a future where a potential TOE may address these ultimate questions. His concept of robustness plays a central role in the author's argument and is proposed as a criterion to distinguish between elements of a theory that represent real entities or structures and those that are merely mathematical artifacts. An element of a theory is considered "robust" if it can be discovered, measured, or derived in various independent ways. This means that a robust theoretical element does not only appear in a specific mathematical model but retains its significance even when assumptions or mathematical frameworks change. This provides a crucial distinction from mathematical artifacts: not all mathematical components of a theory are physically significant. The challenge for scientific realism is to consider only those elements as real that are crucial for empirical success. This means not blindly incorporating every mathematical structure of a theory into ontology but making a critical selection.}.
Therefore, AxQFT should strive to make its axioms more flexible to allow models that are physically relevant and empirically testable, as suggested by D.\;Fraser \cite{Fraser2009}. According to her, this could be achieved by introducing "softer" axioms that accommodate more physical phenomena without completely abandoning mathematical rigor. Initial attempts have been made by Rivasseau \cite{Rivasseau2007} and Bender \cite{Bender2007}. Instead of adhering to extremely high mathematical standards, AxQFT should make deliberate concessions where necessary to make models empirically testable. Conversely, CQFT should attempt to integrate certain axiomatic elements into its methodological repertoire to enhance mathematical rigor without losing touch with empirical verifiability.
\\ \\
This can only be achieved through interdisciplinarity with mathematics. However, today's research landscape presents a different picture: departments of accelerator physics and mathematical physics are often strictly separated in institutes and conferences; there is little dialogue between members of the two programs, and it is not believed that they can mutually enrich each other in any conceivable way. "One would expect set theorists to be vitally interested in the implications of renormalization in quantum field theories (...) for the propriety of their very methods would hang in the balance," Maddy (\cite{Maddy1997}, p.\;159) characterizes the relationship between mathematical and theoretical physics as "social isolation." Although this radical assessment has been relativized by \cite{Rickles2013}, this article is ultimately also an invitation to representatives of both fields to approach each other substantively, to overcome the previous separation, and to transform QFT back into a unified program as in the early 1950s.

The model introduced in this article as a guiding principle for this endeavor may lead to a middle ground that balances the realism quotient of BSM physics in a way that enables long-term theoretical and experimental successes\footnote{Particular emphasis is placed on the term "long-term": the Millennium Problem of a realistic AxQFT is extremely ambitious, even in a weakened form, so that even progress within the 21st century may be considered optimistic. Similarly, patience is required on the experimental front: the long-term limited reach of accelerator experiments is highlighted, whose principal epistemic limits may even indicate the threat scenario of the "end of the particle era." \cite{Harlander2023}}.

\section{Conclusion}
This paper has introduced the realism quotient as a dynamic conceptual framework for navigating the longstanding debate between (mathematical) realism and instrumentalism in the foundations of physics. Our model suggests a dynamic balance of elements from both worlds as a key to the success of theories in the course of experimental progress. Instead of strictly classifying theories as either realistic or instrumentalist, the realism quotient allows for a gradual assessment of the extent to which a theory relies on realistic or instrumental assumptions. It fosters an understanding that theories that are strongly instrumentalist at a given time may later be supplemented or replaced by more realistic ones. The realism quotient provides a foundation for a pragmatic approach to realism, recognizing that future successful theories should both offer increasingly truthful descriptions of reality (enhancing realistic elements) and serve as useful tools for empirical validation (allowing for a significant degree of instrumental elements). Throughout the history of modern physics and technological development, experiments have implicitly sought an optimal quotient that combines the strengths of both positions. The concept of the realism quotient thus enables a rational reconstruction of theory development while also carrying normative force through the extrapolation of its numerical progression. As an example, the model presents an option alongside other recent articles aiming to resolve the poles of the QFT debate. While both QFT programs operate in isolation, our concept instead advocates for a middle ground, in which, at every point in the history of science, a new balancing of realism and instrumentalism is necessary for empirical adequacy, along with the identification of appropriate mathematical structures suited to the current maturity of the discipline. Our concept thus offers a conciliatory perspective in the Fraser-Wallace dispute over the diverging QFT programs while simultaneously revealing its normative potential By proposing the realism quotient as a conceptual compass, it offers a structured yet dynamic guide for future theoretical advancements. In short, reason for introducing the realism quotient model and integrating it into debates on the philosophy of science is twofold:
\begin{enumerate}
\item A return to simplicity: One side-effect of the model is its return to the fundamental yet powerful dichotomy of scientific realism versus instrumentalism. While contemporary philosophy of science has refined and complicated these concepts  (among others, an epistemic/ontic structural, effective, selective, pragmatic, semi-, ...realism), especially when aiming to interpret QFT, our approach suggests that these refinements  may ultimately collapse into a single, more straightforward framework (limited to mathematical structures and thus, foundational physics). By conceptualizing realism as dynamic quotient in scientific history rather than a fixed stance, this model unifies many perspectives within a structured yet flexible design. In addition to its immediate implications for QFT, the realism quotient may fuel broader discussions on the evolution of scientific theories, thereby extending  the model beyond physics and into the general philosophy of scientific progress. Its simplicity ultimately lies in its mathematicity, too, which allows for visualization and extrapolation, offering an intuitive and systematic method for understanding theoretical shifts in physics.
\item A normative potential: Speaking of extrapolation, the realism quotient offers normative guidance for addressing the fragmentation between CQFT and AxQFT - beyond its descriptive power of a rational reconstruction of modern fundamental physics. It thus adds a new dimension for analysing the crisis of physics beyond the Standard Model, while usually the debate centers more around the limited reach of collider experiments and the usage of aesthetics and other metatheoretical criteria like naturalness in theory building. Moreover, it offers a conciliatory perspective in the Fraser-Wallace dispute in philosophy of science. As will be outlined in a future article, out model can be easily applied to competing theories of gravity, offering a conceptual tool for assessing their empirical and mathematical viability in a unified manner.
\end{enumerate}

\subsection*{Acknowledgements}
J.B. wants to thank Michela Massimi, Andreas Hüttemann, Ulrich Krohs, Johannes Heinle and Florian Fabry for valuable discussions and for a critical review. He also expresses his gratitude to the University of Edinburgh, where this article was finalized, for the hospitality. 
\subsection*{Author contributions}
J.B. was the only author of this article
\subsection*{Conflict of interests/funding}
The author states that there is no conflict of interest. He did not receive support from any organization for the submitted work.
\subsection*{Data availability}
No datasets were generated or analysed during the current study.

\end{document}